\documentclass[conference]{IEEEtran}
\def\IEEEtitletopspace{18pt}
\IEEEoverridecommandlockouts
\usepackage{cite}
\usepackage{amsmath,amssymb,amsfonts}
\usepackage{algorithmic}
\usepackage{graphicx}
\usepackage{textcomp}
\usepackage{xcolor}
\usepackage{url} 
\def\BibTeX{{\rm B\kern-.05em{\sc i\kern-.025em b}\kern-.08em
    T\kern-.1667em\lower.7ex\hbox{E}\kern-.125emX}}
\begin{document}

\title{Eye-Tracking and BCI Integration for Assistive Communication in Locked-In Syndrome: Pilot Study with Healthy Participants
}


\author{Ana Patrícia Pinto$^{1}$, Rute Bettencourt$^{2}$, Urbano J. Nunes$^{3}$, Gabriel Pires$^{4}$ 
\thanks{This work was supported by the Portuguese Foundation for Science and Technology (FCT) through the BCI4ALL R\&D Project (COMPETE2030-FEDER-00842800, 2023.17977.ICDT) and by the R\&D Unit UID/00048, Instituto de Sistemas e Robótica – Coimbra (ISR-UC).}
\thanks{$^{1}$Ana Patrícia Pinto  is with the Institute of Systems and Robotics, University of Coimbra, Portugal
        {\tt\small anapinto20851@gmail.com}}%
\thanks{$^{2}$Rute Bettencourt is with Institute of Systems and Robotics, University of Coimbra, Portugal
        {\tt\small rute.bettencourt@isr.uc.pt }}%
\thanks{$^{3}$Urbano J. Nunes is with Institute of Systems and Robotics, University of Coimbra, Portugal, and also with Department of Electrical and Computer Engineering, University of Coimbra, Portugal
        {\tt\small urbano@deec.uc.pt}}%
\thanks{$^{4}$Gabriel Pires is with the Institute of Systems and Robotics, University of Coimbra, Portugal, and also with the Engineering Department of the Polytechnic Institute of Tomar, Portugal
        {\tt\small gpires@isr.uc.pt}}%
}

\maketitle
\enlargethispage{-0.5in}

\begin{abstract}
Patients with Amyotrophic Lateral Sclerosis (ALS) progressively lose voluntary motor control, often leading to a Locked-In State (LIS), or in severe cases, a Completely Locked-in State (CLIS). Eye-tracking (ET) systems are common communication tools in early LIS but become ineffective as oculomotor function declines. EEG-based Brain–Computer Interfaces (BCIs) offer a non-muscular communication alternative, but delayed adoption may reduce performance due to diminished goal-directed thinking. This study presents a preliminary hybrid BCI framework combining ET and BCI to support a gradual transition between modalities. A group of five healthy participants tested a modified P300-based BCI. Gaze and EEG data were processed in real time, and an ET-BCI fusion algorithm was proposed to enhance detection of user intention. Results indicate that combining both modalities may maintain high accuracy and offers insights on how to potentially improve communication continuity for patients transitioning from LIS to CLIS.
\end{abstract}


\section{Introduction}

Communication is a challenge for patients in the Locked-In State (LIS) derived from amyotrophic lateral sclerosis (ALS). These patients lose the ability of spoken communication due to the neurodegenerative nature of ALS, as well as the ability to perform voluntarily controlled movements \cite{Bauer1979,Schnetzer2023}. Eye movement is typically the last voluntary function to be preserved, making eye-tracking (ET) a widely adopted form of alternative and augmentative communication (AAC) for individuals with ALS. However, once eye movements start to fade, leading patients to progress to a Complete Locked-In State (CLIS), the ET systems become unreliable as a communication tool.

Brain-computer interfaces (BCI) use brain signals, commonly acquired with non-invasive techniques, such as electroencephalography (EEG), to enable communication with a computer or device without relying on the usual neuromuscular pathways \cite{Birbaumer2006}. BCIs can be used as an alternative communication method (e.g., P300-speller) and achieve good accuracy with LIS patients. Nonetheless, because ET systems are generally faster and easier to operate, require minimal setup, and provide more intuitive control, patients often continue to rely on them for as long as possible. Patients and caregivers typically consider transitioning to BCI solutions only when eye movement has deteriorated to the point that ET becomes ineffective. This delay often results in a significant communication gap during which the patient is unable to interact effectively. According to the extinction of goal-directed thinking hypothesis \cite{Birbaumer2013}, this prolonged period without effective communication may contribute to the low and inconsistent BCI performance among CLIS patients. This hypothesis suggests that the lack of successful intentional interaction leads to a gradual deterioration of goal-directed cognitive processes, affecting the ability to use BCIs. Therefore, it is crucial to ensure that patients maintain some form of communication and can seamlessly transition between interfaces.

ET is widely used in cognitive psychology and neuroscience, with various AAC solutions ranging from low-tech methods like eye blinks and communication boards \cite{Ezzat2023} to high-tech approaches such as electro-oculography (EOG) and videooculography (VOG). Low-tech solutions, while effective, rely on message interpretation by the receiver, increasing cognitive load and limiting patient autonomy \cite{Voity2024}. High-tech methods address these issues. EOG measures electrical potential differences generated by eye movements, enabling the detection of horizontal and vertical saccades, as well as blinks. Its main disadvantages are its intrusiveness, and the requirement for exaggerated unnatural eye movements to ensure reliable saccade detection. EOG has been explored in assistive applications for motor-disabled individuals, including wheelchair control \cite{Garrote2019, Perdiz2018} and virtual keyboard typing \cite{Usakli2010}. VOG, which uses infrared cameras, measures parameters like pupil position and gaze direction. It is widely used in research and assistive communication for patients with severe motor impairments. Systems such as Tobii Dynavox  \cite{Szymkowicz2024} have significantly improved communication efficiency in this population. In addition, VOG allows the measurement of cognitive workload through metrics like pupil dilation, blink rate, and fixation behavior. Studies have shown correlations between these ocular metrics and task complexity \cite{Escalona2024}, \cite{Tobii2024}, suggesting that workload information extracted from ET could be integrated into hybrid BCI-ET systems to assess user state and potentially optimize system responsiveness.


Integrating BCIs with ET has been explored but remains relatively underdeveloped. In \cite{Ma2018}, a hybrid BCI model combining EEG and VOG ET data was proposed for high-speed text entry. The system combines steady state visual evoked potentials (SSVEP) and gaze proximity, dynamically adjusting decision-making based on classification accuracy of each modality. This approach outperformed single-modality systems in both accuracy and speed. Similarly, hybrid P300 BCI spellers integrating EEG with other modalities, such as VOG \cite{Kalika2017} and EOG \cite{Perdiz2019}, have also demonstrated improved metrics, including accuracy, bit-rate, and error correction.

In this work, we investigate the integration of ET and EEG-based BCI, with the goal of creating a seamless transition strategy for patients with LIS and those possibly progressing toward CLIS. By enabling early adaptation to BCI use, tailoring the system to patient-specific needs, and reducing communication gaps, this approach aims to preserve cognitive engagement and improve long-term outcomes. Additionally, we analyze pupil dilation from ET data to assess user attentional focus during interface control. The approaches were validated with a control group of five able-bodied participants.

\section{Materials and Methods}

\subsection{Participants}

Five healthy female volunteers (22.8 $\pm$ 1.8 years) with normal vision participated in this study. Informed consent was given by all participants.

\subsection{EEG data acquisition}

EEG signal was recorded using the g.USBamp acquisition system (g.tec medical engineering GmbH, Austria) at 256 Hz sample rate, from 16 g.Ladybird electrodes, placed in Fz, Cz, C3, C4, Cz, CPz, Pz, P3, P4, PO7, PO8, POz, Oz, FPz, FCz, FC1, and FC2, according to the extended international 10-20 system. The ground electrode was placed at AFz and the reference at the right earlobe. Signals were filtered with a band-pass filter between 0.1 and 30 Hz and a notch filter at 50 Hz to eliminate powerline interference. EEG signals were acquired, processed, and classified in real-time in a highspeed online processing Simulink framework.

\subsection{ET data acquisition}

In parallel, gaze data and pupil diameter were recorded using the Tobii Pro Spark ET, at a sampling rate of 60 Hz. Data acquisition and real-time processing were carried out using the Titta toolbox \cite{niehorster2020titta}, an open source toolbox that provides a wrap-around of the Tobii Pro SDK. The ET was placed below the acquisition laptop screen. Head position and viewing distance were checked before each session, with participants seated approximately 60 cm from the screen. 

\subsection{ET-BCI Paradigm}

%

The visual BCI interface was adapted from the P300-based Visual Standard Face (Vsf) paradigm\footnote{See demonstrative videos at \url{https://home.isr.uc.pt/~gpires/videos/BCI4ALL/videos.html}} used in \cite{Bettencourt2024} to integrate ET classification. It consisted of a two-row by four-column grid where the seven Portuguese words ‘Sim’, ‘Não’, ‘Tosse’, ‘Ajuda’, ‘Stop’, 'TV', and ‘Sono’ were presented following  an oddball paradigm, and overlaid with a neural face during each stimulus event, using the Psychophysics toolbox \cite{Pelli1997}. Each participant started with a five-point ET calibration. Participants were then instructed to focus on the word displayed at the beginning of each trial (the target) and mentally count the number of times the neutral face flashed over that word, while ignoring the remaining words (non-targets). Each online session, during which participants received visual feedback from the BCI classification, was preceded by a training session in which EEG data were collected and labeled to train the classification model.

\subsection{ET-BCI Fusion Algorithm}
\subsubsection{ET and BCI confidence scores} 
ET data were continuously collected during the P300-oddball stimulus presentation. At the end of each trial, the participant’s gaze coordinates were retrieved in real time using the Titta toolbox. This ensured uninterrupted acquisition of both gaze and EEG data, while allowing efficient access to the ET buffer, storing gaze coordinates, without interfering with the ongoing data streams.
Confidence scores, $C_{\text{ET}}(k)$, for each Area of Interest (AOI) \(k\),  representing both the target and non-target regions, were determined by calculating the ratio of gaze points falling within each AOI to the total number of gaze points recorded for that trial:

\begin{equation}
C_{\text{ET}}(k) = \frac{n_k}{N_{\text{total}}} .
\label{eq:et_equation}
\end{equation}%
The variable \(k\) (\( k\) = 1, ...7) indexes  the AOI of each one of the seven words presented, \( n_k \) is the total number of gaze points within the \( k^{\text{th}} \) AOI, and \( N_{\text{total}} \) is the total number of gaze points collected during the trial. Each gaze point corresponds to the average of the collected left and right gaze coordinates.

In addition, the gaze centroid was also calculated after each trial to monitor the participant’s focus. This allowed real-time assessment of gaze behavior and provided indirect cues about participant's understanding of the task instructions.  

EEG data were decoded using our well established pipeline \cite{Bettencourt2024}, which concludes with a Gaussian Naïve Bayes classifier. For each event, the Bayesian score was computed by combining the prior probability and the class-conditional likelihood:%
\begin{equation}
\text{Score}^k_i = \text{Prior}_i^\top \times \text{pdf}_{\text{Gaussian}}(x \mid \mu_i, \Sigma_i)
\label{eq:bayesian_score}
\end{equation}%
where $\mu_i$ and $\Sigma_i$ are the mean and covariance matrix for class $i$ in the binary (target vs. non-target) classification model, with $i$ = 1, 2. The class label for each event was determined by selecting the class with the highest Bayesian score. The confidence of being a target event was computed for each event from:%

\begin{equation}
\text{C}_{\text{EEG}}^{k} = 1 - \frac{\max(\text{Score}^k_1, \text{Score}^k_2)}{\text{Score}^k_1 + \text{Score}^k_2}
\label{eq:confidence}
\end{equation}%

\subsubsection{BCI and ET fusion}

\begin{figure}[t]
    \centering
    \includegraphics[width=\linewidth]{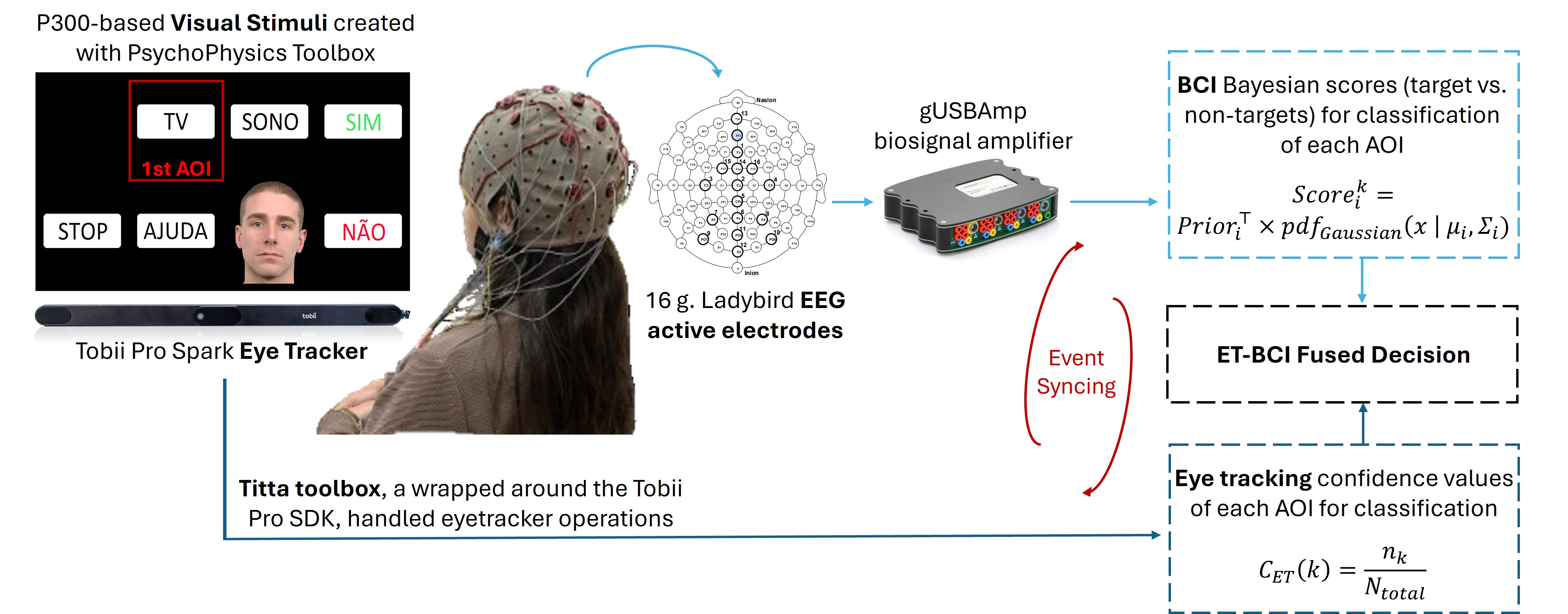}
    \caption{Pipeline illustrating acquisition and classification fusion process of ET and BCI. The system processes gaze and EEG signals independently, extracts confidence scores for each class, and applies a threshold-based decision rule (as defined in (\ref{eq:fused_decision})) to determine the final class prediction.}
    \label{fig:pipeline}
\end{figure}

A hybrid classification approach was implemented by combining ET confidence scores with BCI classifier scores. For each trial, the highest BCI score was identified, and the corresponding ET score was compared to a predefined threshold. The predicted class $\hat{y}(t)$, for trial $t$, was computed as:

\begin{equation}
\hat{y}(t) = \arg \max_{k} \left( \text{C}_{\text{EEG}}^{k}(t) \mid \text{C}_{\text{ET}}^{k}(t) \geq \text{threshold} \right)
\label{eq:fused_decision}
\end{equation}
where $\text{C}_{\text{EEG}}^{k}(t)$ is the BCI classifier score for class $k$, and $\text{C}_{\text{ET}}^{k}(t)$ is the ET score for class $k$. The algorithm first identifies the class with the highest BCI score and checks if the corresponding ET score exceeds the threshold. If the ET score is sufficient, the class is selected. If not, the algorithm moves to the next highest BCI score and repeats the process until a valid class with a sufficient ET score is found. The entire system pipeline is presented in Fig.~\ref{fig:pipeline}.

\section{Results and Discussion}

\subsubsection{Gaze Dispersion}
\begin{figure}[t]
    \centering
    \includegraphics[width=\linewidth]{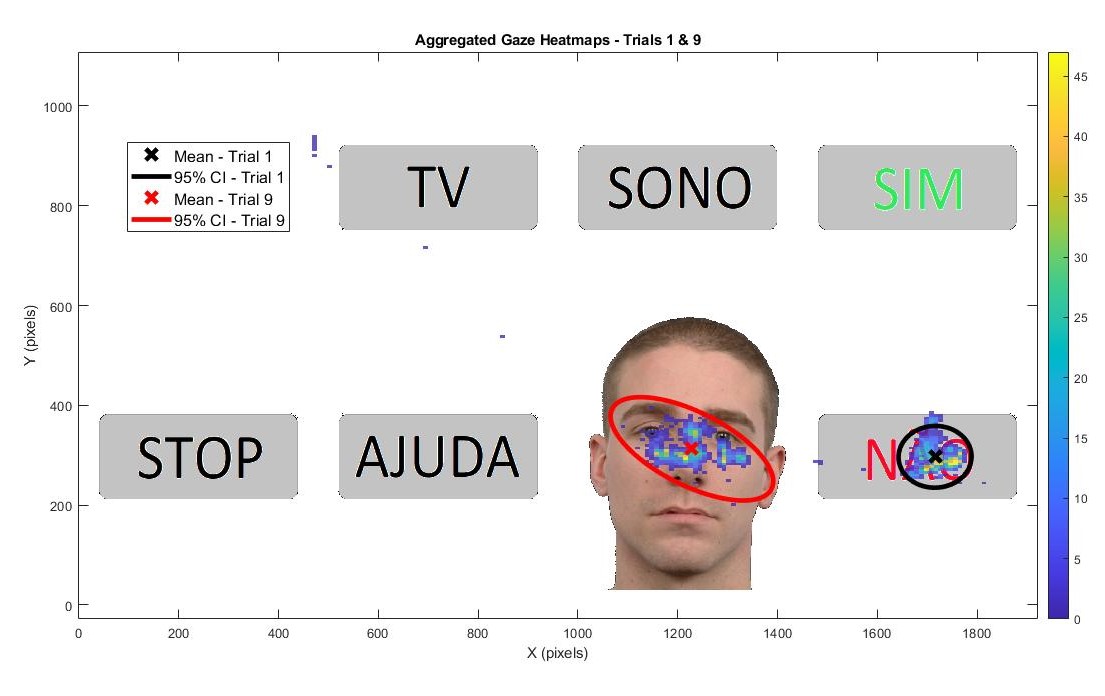}
    \caption{Heatmaps of aggregated gaze overlaid on the BCI visual interface, combining data from all five participants for Trial 1 (black) and Trial 9 (red). The “X” indicates the centroid of the gaze data, and the ellipse denotes the 95\% confidence region, representing spatial dispersion and consistency of visual attention.}
    \label{fig:heatmaps}
\end{figure}

Across all participants and trials, ET data revealed consistent fixation behavior centered around the intended targets, with gaze centroids closely aligned with the corresponding AOIs, as illustrated in Fig.~\ref{fig:heatmaps} for representative trials. 

The 95\% confidence interval ellipsoids offer additional insights into participants' attentional focus. Larger ellipsoid areas, indicating greater dispersion, reflect reduced attentional precision. The average ellipsoid area considering all participants increased from 3928 px² (SD = 852) at the beginning of the session to 4661 px² (SD = 2026) by the final trial, suggesting a possible decline in spatial focus or an increase in cognitive load over time. The rising standard deviation may further indicate greater variability in gaze patterns across participants. The consistent fixation behavior around intended targets indicates that participants correctly understood the task, eliminating the need for explicit verbal confirmation.

\subsubsection{\color{black}ET and BCI Scores}

\begin{figure}[t]
    \centering
    \includegraphics[width=\linewidth]{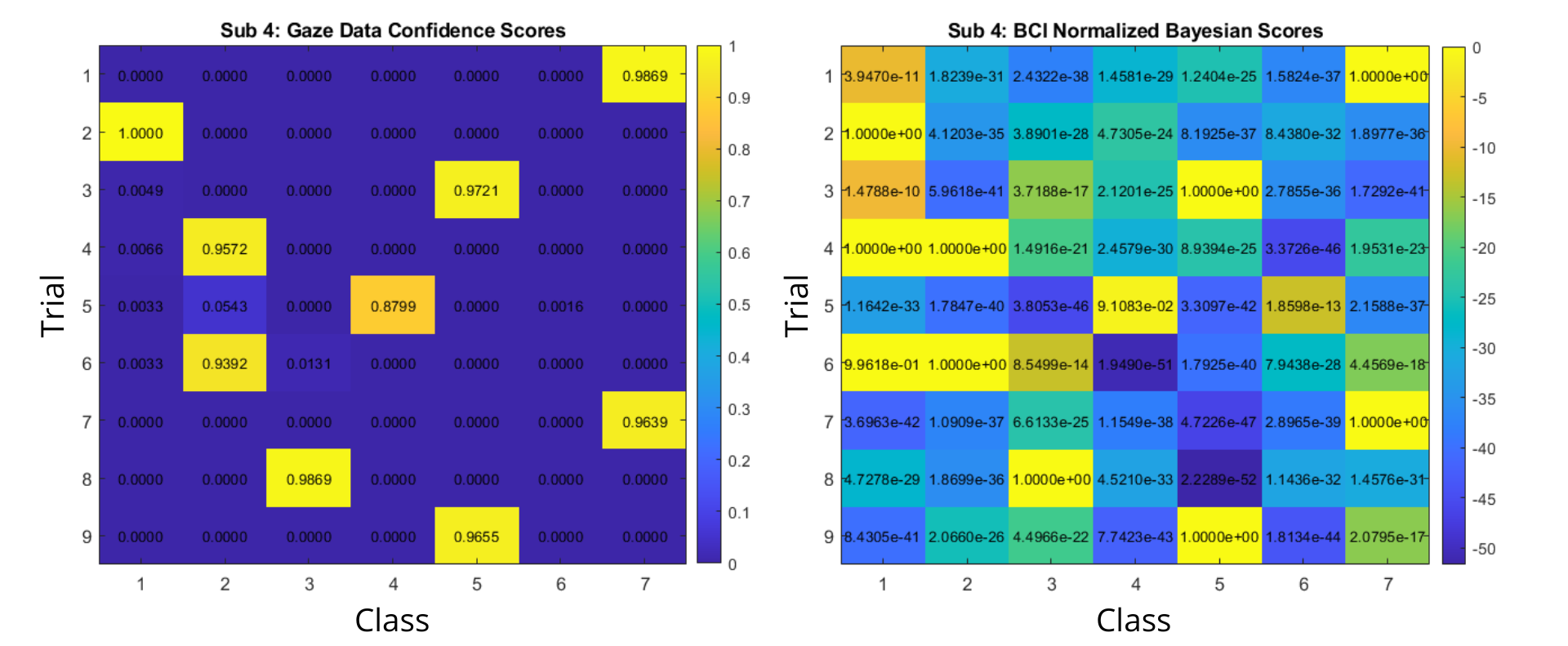}
    \caption{Colormaps representing ET confidence scores (left), determined using (\ref{eq:fused_decision}), and normalized Bayesian BCI scores (right), determined using (\ref{eq:bayesian_score}). Data from Subject 4 across 9 trials and 7 AOIs. The BCI score colormap uses a \(\log_{10}\) scale to enhance visual contrast and highlight relative differences.}
    \label{fig:scores}
\end{figure}

Fig.~\ref{fig:scores} illustrates the confidence scores of ET gaze data and normalized Bayesian scores for the BCI across 9 trials and 7 AOIs for Subject 4. The left colormap provides a visual representation of the distribution of gaze points, indicating where the participant was looking during each trial, while the right colormap indicates BCI confidence in classifying each AOI as the trial’s target class (in comparison to all non-target classes).

In the ET confidence colormap (left), gaze pattern was generally focused during early trials, with a slight increase in variability observed in later trials, which may indicate reduced attentional focus. This increase in dispersion was expected to be subtle, given that the control group consisted of healthy individuals without vision or motor impairments. However, it is important to note that such variations would likely become more pronounced in populations with motor impairments affecting eye movements (e.g., LIS), or when using interfaces with smaller AOIs (e.g., speller in \cite{Usakli2010}).

In parallel, the BCI confidence scores (right) reveal high confidence in classifying the correct target during the initial trials, which aligns with the generally higher ET confidence scores observed at the beginning. However, in trials such as 5 and 6, where ET confidence is lower, a corresponding variation in BCI scores is evident, potentially reflecting shifts in participant engagement or focus.

\subsubsection{Joint Prediction Method}
Both the BCI and ET classification methods demonstrated high accuracy, with average classification accuracies exceeding 90\% and standard deviations below 9.5\%. Although these results align with expectations for a control group of healthy participants, the true potential emerges from the possibility of integrating both modalities. When the fusion threshold used for decision was experimentally set to 0.85 (in (\ref{eq:fused_decision})), all trials were correctly predicted, except for trial 8 of subject 1. In this case, the subject's attention was diverted, causing the participant to focus on the incorrect word. All three classification modalities predicted the same incorrect class, confirming that the error was due to diverted attention rather than a failure of the classification methods.

\subsubsection{Pupil Dilation}

\begin{figure}[t]
    \centering
    \includegraphics[width=\linewidth]{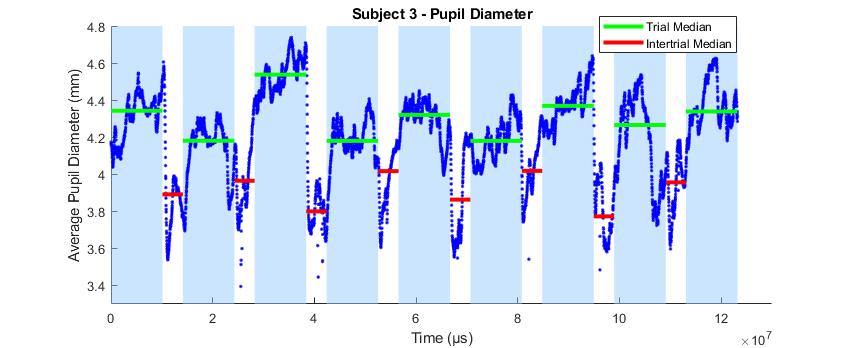} 
    \caption{\color{black}Average pupil diameter during trial and inter-trial periods for Subject 3. Light blue shaded areas represent trial periods.}
    \label{fig:pupil_plot}
\end{figure}

Pupil diameter was recorded as a secondary measure of cognitive effort and arousal throughout each session. Across all participants, pupil diameter was consistently larger during trial periods compared to inter-trial periods, as illustrated in Fig.~\ref{fig:pupil_plot} for Subject 3. This pattern is commonly observed in cognitive load studies and reflects an increase in mental effort.

The median pupil diameter during trial periods across all subjects was 3.2945 mm (SD = 0.738), while during inter-trial periods, it was smaller at 3.1552 mm (SD = 0.525). This observation aligns with existing literature, which links increased pupil dilation to higher cognitive load, mental effort, decision-making, and attentional demand \cite{Tobii2024}.
These preliminary findings indicate that pupil dilation has potential as an indicator of cognitive load and participant engagement, offering further insights into the temporal dynamics of mental effort throughout task execution.

\section{Conclusion}

This preliminary study investigated the integration of ET and EEG-based BCI in a P300-based visual paradigm, aiming to support a smooth transition for patients progressing from LIS to CLIS. Results with healthy participants showed high classification accuracy, and the ET-BCI fusion demonstrated potential for improving real-time performance. This approach lays the groundwork for a gradual adaptation to BCI-only use, enabling continuous patients' communication as motor control declines. The ET data revealed consistent gaze patterns, while pupil dilation offered valuable insights into cognitive load, further enhancing system responsiveness. Future work will include tests with a larger and more diverse control group, extend the protocol to LIS patients, and incorporate more complex BCI interfaces, such as spellers with smaller AOIs, to further validate the fusion strategy.






\bibliographystyle{IEEEtran}
\bibliography{refs.bib}
%


\end{document}